# Effect of partial substitution of iso-valent Mo at Cr-site on electronic structure and physical properties of Fe$_2$CrAl

Kavita Yadav and K. Mukherjee

School of Basic Sciences, Indian Institute of Technology, Mandi, Himachal Pradesh-175005, India

**Abstract**

Heusler alloy Fe$_2$CrAl exhibits a ferromagnetic behaviour below Curie temperature ($T_C$) ~ 202 K along with presence of cluster glass (CG) phase near freezing temperature ($T_f$) ~ 3.9 K and Griffiths phase (GP) above 300 K. The physical properties of this alloy are very sensitive to substitutions and anti-site disorder. Here, we investigate the effect of partial substitution of Mo at Cr-site on physical properties of Fe$_2$CrAl. Structural and morphological analysis confirms the single phase cubic structure of the substituted alloys. Increment in Mo concentration shifts the $T_C$ towards lower temperature, which is ascribed to the effect of increased hybridization strength between 3$d$-4$d$ states of Fe/Cr/Mo. Additionally, systematic analysis of AC susceptibility, magnetic memory effect and time dependent magnetization studies confirm the presence of CG-like phase near ($T_f$) ~ 3.5 K in Fe$_2$Cr$_{0.95}$Mo$_{0.05}$Al. Such feature gets suppressed towards lower temperature with an increase of Mo concentration, i.e. below 1.8 K in Fe$_2$Cr$_{0.85}$Mo$_{0.15}$Al. The origin of the glassy signature is ascribed to the decrement in magnetic anisotropy with Mo concentration. A partial increment in magnetic entropy change is also noted near $T_C$ with the increase in Mo substitution. Interestingly, at high temperatures (above 350 K), GP phase persists in both the alloys due to the presence of anti-site disorder.

## 1. Introduction

In past few decades, Heusler alloys have gained considerable attention and have been extensively investigated as they exhibit unusual physical properties like, unconventional superconductivity [1], half-metallicity [2], magneto–optical effect [3], magnetocaloric effect [4], spin/cluster glass phase [5] etc. In general, the ternary full Heusler alloys are represented by $X_2YZ$, where X and Y are transition metals and Z is *sp* element belonging to group III-VI of periodic table. Ideally, Heusler alloys crystallize in $L2_1$ structure which is the most ordered state. Prediction of half-metallicity has renewed the research interest in such alloys. In general, half metallic materials are characterized by simultaneous presence of energy band gap in one of the sub spin bands at Fermi level ($E_F$) and continuous band in other [2, 6-8]. As a consequence, the conduction electrons present at Fermi level are 100% spin polarized, resulting in unusual electrical properties useful for spintronic applications. The magnetocaloric effect (MCE) is also another fascinating property found in Heusler alloys [4]. Ni–Mn–Z (Z = In, Sn, Sb) shape memory Heusler alloys, which exhibit first order phase transition (FOPT), have been widely studied for the MCE [9-12]. The peak value of $\Delta S_M$ is found to be very large in FOPT alloys, which is required for better cooling efficiency. But due to presence of thermal and magnetic hysteresis, their use in refrigeration applications is limited. This also motivates researchers to search for alternative members in the Heusler alloys family.

Apart from MCE and half-metallicity, these alloys exhibit exotic glassy magnetic phases. This behaviour generally arises due to competition between ferromagnetic (FM) and antiferromagnetic (AFM) correlations arising due to anti-site disorder. Spin glass phase was observed in $Cu_2VAl$ where the structural disorder between V and Al atoms breaks the long range magnetic ordering and results in the formation of individual spins or V clusters [5]. Similar type of glassy phase had been also observed in $Fe_2TiSn$, $Fe_2VAl$ and $Fe_2VGa$ [13]. Through various experimental investigations, it had been also observed that physical properties of Heusler alloys are sensitive towards elemental substitution and anti-site disorder [14-18]. For example, $Fe_2VAl$ exhibits non-metallic behaviour despite metallic constituents and is also reported to be non-magnetic despite of Fe content [19]. Also, on other hand $Fe_2TiAl$ displays metallic character but is weakly magnetic [8]. One of the alloys of this family is $Fe_2CrAl$. Our previous studies on this alloy suggest that it stabilizes in $L2_1$ structure, showing FM behaviour below Curie temperature ($T_C$) ~ 202 K along with cluster glass (CG) phase near 3.9 K [20]. Although from band structure calculations, this system was predicted as half metallic ferromagnetic (HMF) [21], but experimentally it behaves as semiconductor with higher density of states compared to $Fe_2VAl$ [19]. $Co_2$, $Fe_2$, $Mn_2$ based full ternary systems in which $4d$ element is present at X-site are well explored. These investigations reflect the role of $4d$ elements in influencing the electronic structure and magnetic behaviour of full ternary alloys. However, the effect of presence of $4d$ element at Y-site has not been explored from viewpoint of magnetism. Therefore, it will be interesting to investigate the $Fe_2$ based ternary alloys with presence of one magnetic element with $4d$ electrons at Y-site. Hence, this study will help us to understand how the partial presence of Mo ($4d$ element) in place of Cr ($3d$ element) influences the electronic structure and physical properties of $Fe_2CrAl$.

With these objectives, in this manuscript, we present the results of our investigation of the electronic structure and physical properties of $Fe_2Cr_{0.95}Mo_{0.05}Al$ and $Fe_2Cr_{0.85}Mo_{0.15}Al$. Structural analysis reveals that both alloys crystallize in cubic structure. From magnetization results, it is revealed that $Fe_2Cr_{0.95}Mo_{0.05}Al$ undergoes paramagnetic (PM) to FM phase transition near $T_C \sim 190$ K, which is shifted to 160 K in $Fe_2Cr_{0.85}Mo_{0.15}Al$. These values are smaller as observed in $Fe_2CrAl$ and this can be ascribed as the effect of increased hybridization between $3d$-$4d$ states of Fe/Cr/Mo with Mo substitution. An increment in magnetic entropy change is noted with an increase in Mo concentration at Cr-site in $Fe_2CrAl$. Similar to parent alloy cluster glass-like phase is noted near the freezing temperature ($T_f$) ~ 3.5 K in $Fe_2Cr_{0.95}Mo_{0.05}Al$, which is shifted below 1.8 K in $Fe_2Cr_{0.85}Mo_{0.15}Al$. This behaviour can be attributed to the effect of magnetic anisotropy, which decreases with an increase in Mo concentration. Due to the presence of anti-site disorder, Griffiths phase (GP) persists in PM region of both alloys.

2. **Experimental procedure**

High purity elements Fe (≥99.9%), Cr (≥ 99.9%), Mo (≥ 99.9%) and Al (≥ 99.9%) are taken in stoichiometric ratio to synthesize the polycrystalline alloys of $Fe_2Cr_{1-X}Mo_XAl$ (x = 0.05 and 0.15) using arc melting technique. The weight loss is found to be less than 0.2 % (in both cases) signifying that the nominal composition assigned to each alloy are accurate. The resultant ingots are subjected to annealing at 1223 K for 3 weeks in vacuum sealed quartz ampoules (>$10^{-5}$ mbar). The annealing is employed for better homogenization and to avoid formation of by-products. The alloy $Fe_2CrAl$ used for investigation of electronic structure and magnetocaloric properties is the same as mentioned in Ref. 20. To confirm the phase purity and crystal structure of alloys, powder X-ray diffraction (XRD) patterns at room temperature are obtained using rotating anode Rigaku Smartlab diffractometer in Bragg-Brentano geometry (Cu-Kα; λ=1.5406 Å). Morphological and elemental analysis of alloys is carried out by employing field emission scanning electron microscope (FESEM, JFEI, USA, Nova Nano SEM). Room temperature X-ray photoelectron spectroscopy (XPS) measurements of alloys are done on a NEXSA surface analysis model by Thermo Fisher scientific. Magnetic property measurement system (MPMS), Quantum Design, U.S.A is used to perform temperature and field dependent DC and AC susceptibility measurements.

## 3. Results and discussions

### 3.1 Structural and Morphological analysis-

The formation of single phase $Fe_2Cr_{1-x}Mo_xAl$ (x = 0.05 and 0.15) alloys with cubic structure is confirmed with the indexed XRD patterns (Fig. 1). The lattice parameter of $Fe_2CrAl$ is reported to be ~ 5.784 Å. With increasing Mo content, a systematic increment in lattice parameter ($a$) is noted with respect to parent alloy, which is in accordance with Vegard's law. The obtained parameters are tabulated in Table 1. As noted from table 1, the increment in '$a$' is very small in $Fe_2Cr_{0.95}Mo_{0.05}Al$, hence, no significant shift in (200) peak position is observed (as shown in inset of Fig. 1). However, this shift is visible in $Fe_2Cr_{0.85}Mo_{0.15}Al$ alloy. As noted from Fig. 1, the Bragg peaks corresponds to the (220), (400) and (422) reflections of cubic unit cell. The absence of the (200) and (111) reflections indicate towards the formation of a disordered structure. This disorder can influence the physical properties and electronic structure of Heusler alloys, as they are sensitive towards structure and composition; minor changes in reflection peaks and its intensity can also have impact on these properties [22-26]. In fact our previous studies on $Fe_{2-x}Mn_xCrAl$ (0≤x≤1) series reveal that a decrement in the intensity ratio $I_{200}/I_{220}$; with an increment of Mn substitution at Fe-site [20]. This decrement results in an increase in anti-site disorder with Mn substitution. Morphological analysis is performed on pellets of $Fe_2Cr_{0.95}Mo_{0.05}Al$ and $Fe_2Cr_{0.85}Mo_{0.15}Al$ and representative SEM images of the polished surfaces are shown in Fig. 2 (a) and (g) respectively. As seen from the figure, both the alloys have clean surface without the presence of any secondary phase/patches. Further, elemental mapping for Fe, Cr, Mo and Al (Fig. 2 (b)-(f) and Fig. 2(h)-(i)) depict that all the elements are uniformly distributed within the alloys. These observations are consistent with XRD data and reflect the homogeneity of the prepared alloys.

## 3.2 Electronic structure of $Fe_2Cr_{1-x}Mo_xAl$ (x = 0, 0.05 and 0.15)

In order to study the effect of simultaneous presence of 4*d* and 3*d* element on same site on electronic structure of the $Fe_2Cr_{1-x}Mo_xAl$ (x = 0, 0.05 and 0.15), we have performed x-ray photoemission spectroscopy (XPS) at room temperature using a monochromatic Al-K$_\alpha$ x-ray source (hν = 1486.6 eV). Fig. 3 (a), (b) and (c) shows the Fe 2*p*, Cr 2*p*, Al 2*p* core level XPS spectra of Fe$_2$CrAl respectively, measured with same instrumental settings. We have subtracted the in-elastic background with Tougaard method [27] and fitted the core level spectra with Voigt function (it includes instrumental broadening parameters). For Fe$_2$CrAl, we have observed spin-orbit splitting of peaks of Fe 2*p* position near 707.30 eV and 720.21 eV corresponding to 2*p*$_{3/2}$ and 2*p*$_{1/2}$ respectively with spin-orbit splitting energy of about 12.91 eV. These observations are consistent with literature [28]. Here, it can be seen that shape of both branches are significantly different from each other as 1/2 branch has large width compared to 3/2 branch. This feature can be explained in terms of a smaller core-hole lifetime due to availability of super coster-kronig transitions [29]. Also, statistical branching ratio of both branches (2*p*$_{3/2}$ : 2*p*$_{1/2}$) i.e. intensity ratio is found to be 2:1, which is equal to that expected from the (2*j*+1) multiplicity of the states. Additionally, while fitting the 2*p* spectra of Fe, extra components are used to get proper fit. This extra component belongs to multiplet structure of 2*p*$_{3/2}$ and 2*p*$_{1/2}$, respectively. The slight shift in the peak position can be due to the hybridization between Fe and Cr states as well as can be due to interactions between core holes and conduction electrons. Similarly, we have observed Cr 2*p*$_{3/2}$ and 2*p*$_{1/2}$ peaks near 574.46 eV and 583.63 eV. Here, we have noted that Cr 2*p* spectrum do not show any exchange splitting. This suggests the presence of strongly delocalized Cr 3*d* states and strong hybridization between Fe and Cr gives rise to magnetism in this alloy. It has also been reported in other 3*d* based Heusler alloys [30]. From the Al 2*p* spectrum, we have observed two peaks near 72.67 eV and 74.85 eV corresponding to 2*p* and Al$_2$O$_3$ respectively. This oxide peak of Al can originate due to atmospheric exposure of the alloy. Also, as the difference between 2*p*$_{3/2}$ and 2*p*$_{1/2}$ is only 0.4 eV, it is difficult to distinguish between 2*p*$_{3/2}$ and 2*p*$_{1/2}$ peak in Al 2*p* spectra. Hence, we have fitted the 2*p* peak using three components: 2*p*$_{3/2}$, 2*p*$_{1/2}$ and satellite peaks. The later one is used as we have noted satellite peaks in this spectrum. This peak can be attributed to inter-band transitions and reflect the correlation of the angular momentum of electron core hole with the local band structure. The binding energy value of 2*p*$_{3/2}$ peak corresponding to Cr and Al is slightly large as compared to the binding energy of elemental Fe $2p_{3/2}$ (706.8 eV) and Al $2p_{3/2}$ (72.6 eV), respectively. This is possibly due to the hybridization between partially filled 3$d$ levels.

In $Fe_2Cr_{0.95}Mo_{0.05}Al$ and $Fe_2Cr_{0.85}Mo_{0.15}Al$, we have observed similar behaviour of $2p_{3/2}$ and $2p_{1/2}$ peaks corresponding to Fe 2$p$, Cr 2$p$ and Al 2$p$. The corresponding spectra are shown in Fig. 3 (d)-(k). However, with Mo substitution, chemical shift of these peaks towards higher binding energy side is noted (Table 2). In $Fe_2Cr_{0.95}Mo_{0.05}Al$ alloy, we have also observed spin-orbit splitting peaks of Mo 3$p$ near 394.13 eV and 411.81 eV, which corresponds to the $3p_{3/2}$ and $3p_{1/2}$, respectively. These peaks get shifted to 394.29 eV ($3p_{3/2}$) and 411.83 eV ($3p_{1/2}$) in $Fe_2Cr_{0.85}Mo_{0.15}Al$. Here, strong hybridization is present between the 4$d$ orbital of Mo and 3$d$ orbital of Fe, which is responsible for magnetism in this series of alloys. As 4$d$ orbital is more extended compared to 3$d$ orbital, there are significant chances of strong overlap between 3$d$ and 4$d$ orbital. This is also reflected in increased binding energy of respective elements with Mo substitution as the hybridization between the $d$-orbitals of Fe-Cr/Mo becomes stronger. Hence, along with 3$d$-3$d$ hybridization of Fe-Cr elements, 3$d$-4$d$ hybridization of Fe-Mo elements also plays a crucial role in observed magnetic moment in both alloys, despite of same number of valence electrons in Cr and Mo. Additionally, we have also noted satellite peaks in Cr, Mo, Al spectra of Mo doped alloys.

### 3.3 Temperature and field response of DC magnetic susceptibility

Temperature response of DC susceptibility of the $Fe_2Cr_{0.95}Mo_{0.05}Al$ and $Fe_2Cr_{0.85}Mo_{0.15}Al$ alloys at 100 Oe under zero field cooled (ZFC) and field cooled (FC) conditions are shown in Fig. 4 (b-c). Our earlier report suggest that $Fe_2CrAl$ undergoes PM to FM transition near 202 K ($T_C$) (Fig. 4(a)) [20]. In the present case, it has been observed that increment in Mo substitution at Cr-site reduces this magnetic transition temperature to 191 K and 160 K in $Fe_2Cr_{0.95}Mo_{0.05}Al$ and $Fe_2Cr_{0.85}Mo_{0.15}Al$, respectively. Similar to parent alloy, we have also noted weak irreversibility between ZFC and FC curves below the irreversible temperature ($T_{irr}$) ~ 70 and 50 K respectively, as shown in right insets of Fig. 4 (b-c). Generally, this type of behaviour signifies the presence of glassy phase or can arise due to domain wall motion [31-33]. Additionally, as observed in left inset of Fig. 4 (b)-(c), in high temperature regime of $Fe_2Cr_{0.95}Mo_{0.05}Al$, we have also observed magnetic anomaly above $T^*$ ~ 350 K which is increased to 380 K in $Fe_2Cr_{0.85}Mo_{0.15}Al$. In $Fe_2CrAl$, this feature has been ascribed as presence of GP in the alloy and will be discussed in detail in the later section.

Figure 4 (d-e) shows field dependent isothermal magnetization $M$ ($H$) curves for $Fe_2Cr_{1-x}Mo_xAl$ (x = 0.05 and 0.15) at 2 K and 300 K. At 2 K, in $Fe_2Cr_{0.95}Mo_{0.05}Al$, we have observed a negligible magnetic hysteresis. Additionally, a non-saturating behaviour is noted which reflects the presence of short-ranged correlations along with FM behaviour (which is shown in inset of Fig. 4 (f)). Similar behaviour is also noted in $Fe_2Cr_{0.85}Mo_{0.15}Al$. This implies that both alloys exhibit soft FM behaviour. To investigate the presence of magnetic anisotropy in both alloys, we have analyzed the $M$ ($H$) data at 2 K (in the range $5 \leq H \leq 50$ kOe) using the equation (1). In equation (1), $H_r$ and $H_{ex}$ is the anisotropic and exchange field, respectively [20]. The fitted curves are shown in inset of Fig. 4 (f).

$$M = M_0 \left(1 - \frac{1}{15}\left(\frac{H_r}{(H + H_{ex})^2}\right)\right) \ldots \ldots \ldots (1)$$

The parameters extracted from the fit are given in the table 3. It is noted that both $H_r$ and $H_{ex}$ decreases with Mo concentration which indicates a decrement in strength of magnetic anisotropy as compared with $Fe_2CrAl$.

According to Slater Pauling (S-P) equation [34]: $M_t = N_v - 24$ where $M_t$ and $N_v$ are magnetic moment in Bohr magneton per formula unit ($\mu_B$/f.u.) and number of valence electrons per f.u. respectively. Here, for $Fe_2CrAl$, theoretical value from S-P equation is $M_t = 1$ $\mu_B$/f.u. but from $M$ ($H$) it is calculated to be ~ 1.5 $\mu_B$/f.u. This deviation implies that there is presence of anti-site disorder. It is in accordance with the structural analysis of the alloy as reported in Ref. 20. As Mo and Cr are isovalent, total number of valence electrons in the system should remain the same, even if the concentration of Mo is increased. Therefore, the calculated magnetic moment according to S-P equation should be same as that of parent alloy. But as noted from Fig. 4 (d-e), $M_t$ increases with increment in Mo concentration which signifies the amount of disorder increases with Mo substitution. Additionally, at 300 K, for both alloys, it is seen that $M$ ($H$) exhibits non-linear behaviour instead of expected linear behaviour which signifies the presence of some short-range correlations above $T_C$.

**3.4 Time dependent magnetization study-**

In order to shed some light on the observed irreversibility behaviour in $Fe_2Cr_{0.95}Mo_{0.05}Al$ and $Fe_2Cr_{0.85}Mo_{0.15}Al$, isothermal magnetization measurements (IRM) has been performed at 2, 4 and 10 K in $Fe_2Cr_{0.95}Mo_{0.05}Al$ and 2 and 4 K in $Fe_2Cr_{0.85}Mo_{0.15}Al$. The protocol mentioned in Ref. 20 is followed and the obtained curves are shown in Fig. 5 (a-b). The obtained curves are fitted with-

$$M(t) = M_0 - S\left(\ln\left(1 + \frac{t}{t_0}\right)\right) \ldots\ldots (2)$$

where $M_0$ is the magnetization value at $t = 0$ and $S$ is the magnetic viscosity [31,35]. The obtained parameters are given in table 4. At 2 and 4 K, it is seen that in $Fe_2Cr_{0.95}Mo_{0.05}Al$, magnetization decays with time, whereas, no change in magnetization is noted at 10 K. Similar relaxation behaviour is noted in $Fe_2Cr_{0.85}Mo_{0.15}Al$ at 2 K. However, at 4 K no decay in magnetization is noted. Here, logarithmic decay indicates a distribution of cluster of spins separated by different energy barriers in the system. This indicates the possibility of glassy phase in this series of alloys below 10 K and 4 K in $Fe_2Cr_{0.95}Mo_{0.05}Al$ and $Fe_2Cr_{0.85}Mo_{0.15}Al$, respectively.

**3.5 ZFC memory effect**

Above relaxation measurements indicate the presence of non-equilibrium dynamics in $Fe_2Cr_{0.95}Mo_{0.05}Al$ and $Fe_2Cr_{0.85}Mo_{0.15}Al$ below 10 and 4 K respectively. To further investigate this phase, we have performed ZFC memory effect experiment using the following protocol [36, 37]: the alloys are cooled in presence of zero field, and halts are made at 4 K and 10 K ($Fe_2Cr_{0.95}Mo_{0.05}Al$) and 3 K ($Fe_2Cr_{0.85}Mo_{0.15}Al$) for 2 hrs. Then, 100 Oe is applied and data is obtained during warming cycle (ZFC$_{\_tw}$). The obtained curves are shown in Fig. 5 (c)-(d). Here, in both $Fe_2Cr_{0.95}Mo_{0.05}Al$ and $Fe_2Cr_{0.85}Mo_{0.15}Al$, it is observed that there is significant difference between ZFC$_{\_ref}$ and ZFC$_{\_tw}$ i.e. presence of memory dips near halt temperatures 4 and 3 K respectively. In $Fe_2Cr_{0.95}Mo_{0.05}Al$, at 10 K no memory dip is noted. Generally, memory dips are observed in case of spin glass/CG phase as there is existence of large number of states. As the time progresses, the system goes into deeper valleys with high energy barriers. So, when the temperature is low, the system becomes resistant to applied field causing difference between magnetization with and without stops; which is reflected as memory dips. However, in superparamagnetic (SPM) systems, there is absence of memory dips as the occupation probabilities of up and down spin particles are equal to 0.5 (as seen in two-state model) [36,37]. Thus, in both alloys ZFC memory effect is noted, which indicates the presence of interacting spin clusters in this regime.

### 3.6 AC susceptibility measurements

Results of the previous sections indicate the presence of glassy phase at low temperatures in $Fe_2Cr_{0.95}Mo_{0.05}Al$ and $Fe_2Cr_{0.85}Mo_{0.15}Al$. In order to probe the nature of the underlying glassy magnetic phase in this temperature regime, we have performed the temperature dependent in-phase ($\chi'$) and out-of-phase ($\chi''$) AC susceptibility measurements at different frequencies (as shown in the Fig. 6 (a)-(d)). In $Fe_2Cr_{0.95}Mo_{0.05}Al$, it is observed that $\chi'$ shows frequency dependent broad feature in the low temperature region, while only single frequency dependent peak around 3.5 K is clearly seen in $\chi''$ (inset of Fig. 6 (b)). Similarly, for $Fe_2Cr_{0.85}Mo_{0.15}Al$, it is found that $\chi'$ also exhibit frequency dependent feature but $\chi''$ does not show any well-defined peak, probably due to shifting of $T_f$ below 1.8 K. Here, it can be said that with Mo concentration, $T_f$ shifts towards lower temperature as compared to $Fe_2CrAl$, probably due to decrement in strength of magnetic anisotropy with increment of Mo at Cr-site. Additionally, it can be seen from Fig. 6 (a)-(d) that peaks near $T_C$ in both alloys does not show any shift in temperature with applied frequency as compared to $T_f$. In $Fe_2Cr_{0.95}Mo_{0.05}Al$, it is seen that $T_f$ shifts towards higher temperature side with increment in magnitude of applied frequency. This feature can be due to presence of glass-like freezing phenomenon. In order to identify the nature of this glassy phase around $T_f$, we have determined Mydosh parameter ($\delta T_f$) [31]

$$\delta T_f = \frac{\Delta T_f}{T_f \log f} \ldots\ldots\ldots\ldots (3)$$

where $T_f$ is the peak temperature corresponding to frequency (*f*). Our recent studies on $Fe_2CrAl$ reveal the presence of CG-like phase near $T_f \sim 3.9$ K. In $Fe_2Cr_{0.95}Mo_{0.05}Al$, $\delta T_f$ is found to be nearly 0.154 which lies in CG regime [23, 31-33]. Behaviour of magnetic clusters (interacting/non-interacting) present in this regime is also analysed with power-law and Vogel-Fulcher (VF) laws [31-33]. In case of power-law, the temperature dependence of relaxation time is given by

$$\tau = \tau_0 \left(\frac{T_f - T_g}{T_g}\right)^{-zv} \ldots\ldots\ldots (4)$$

where $\tau_0$ is the microscopic flipping time, $T_g$ is the true spin glass (SG) transition temperature, *z* is the dynamic critical exponent and *ν* is the critical exponent of the correlation length. The values of parameters obtained after fitting (as shown in left inset of Fig. 6 (a)) are $\tau_0 = 9.8 * 10^{-6}$ s and $zv = 2.46 \pm 0.24$. In conventional SG systems, the value of *zv* lies between ~ 2 and 12 while the value of $\tau_0$ lies in the range $10^{-10}$ - $10^{-13}$ s [38, 39]. Similarly, in case of CG system, the range of $\tau_0$ is from $10^{-5}$ - $10^{-10}$ s [20, 36, 38, and 40]. The obtained parameters indicate the presence of CG phase in this alloy. The temperature maxima of $\chi''$ (T) is fitted with V-F law (which takes into account the interaction between spins), best fit is obtained with V-F law of the form

$$\tau = \tau_0 \exp\frac{E_a}{k_B(T_f - T_0)} \ldots\ldots\ldots (5)$$

where $\tau$ is the magnetization flip time of particle/cluster, $\tau_0$ is the attempt time, $E_a$ is the activation energy of the relaxation barrier, $T_f$ is the freezing temperature at each frequency and $T_0$ is the V-F temperature, which represents the strength of interactions between clusters. In $Fe_2Cr_{0.95}Mo_{0.05}Al$, the parameters obtained after fitting (as shown in right inset of Fig. 6 (a)) are $\tau_0 = 5.8 * 10^{-6}$ s and $T_0 = 3.25$ K which are physical in nature and are in analogy with CG systems. This further supports the fact that the dynamics in this system is due to interacting cluster of spins. This presence of CG phase in low temperature of $Fe_2Cr_{0.95}Mo_{0.05}Al$ can be attributed as the effect of magnetic anisotropy [20]. This anisotropy grows larger with decrement in temperature such that it weakens the coupling leading to formation of CG phase in this alloy.

### 3.7 Presence of Griffiths phase in high temperature region

As noted from Fig. 4, in $Fe_2Cr_{0.95}Mo_{0.05}Al$, as temperature is decreased, below $T^*$~350 K, a deviation from PM behaviour is observed. This deviation is noted at $T^*$ ~380 K for $Fe_2Cr_{0.85}Mo_{0.15}Al$. It suggests the presence of another phase in PM region. It is also observed from Fig. 6 that $\chi'$ exhibit frequency independent peak near these anomalies which rules out the presence of SPM-like clusters near $T^*$. Similar behaviour is noted in parent alloy. One of the hallmarks of GP is a downward deviation in inverse DC susceptibility ($\chi^{-1}$) from Curie-Weiss (CW) law well above $T_C$ [41, 42]. Hence, to identify whether the observed feature can be ascribed to GP, inverse $\chi_{DC}$ as a function of temperature at applied magnetic field of 100 Oe is shown in Fig. 7 (a-b) for $Fe_2Cr_{0.95}Mo_{0.05}Al$ and $Fe_2Cr_{0.85}Mo_{0.15}Al$. In the former alloy a downward deviation is observed, and it is also noted that as the field is increased from 0.1 to 1 kOe, there is softening of downturn. This is a signature of GP and has been reported in various magnetic systems [43-45]. Similar feature is also noted in $Fe_2Cr_{0.85}Mo_{0.15}Al$ near 380 K. The above observations indicate the non-analytical nature of DC susceptibility is arising due to Griffiths singularity [46] and has been well explored in manganese based oxides and iridates [39-44]. Generally, Griffiths singularity is characterized using the susceptibility exponent ($\lambda$) obtained using the equation [42]

$$\chi^{-1} \sim (T - T_C^R)^{(1-\lambda)} \ldots\ldots (6)$$

where 0≤λ<1 and $T_C^R$ is the critical temperature of random FM where χ tends to diverge. This relation is the modified form of CW law where λ signifies the deviation from CW behaviour. The value of exponent λ is ideally equal to 0 in pure PM region, but it yields finite value lying between 0 and 1 above $T_C$. Thus, we can say that higher the value of λ, stronger will be the deviation from CW behaviour. In order to estimate the value of λ for $Fe_2Cr_{0.95}Mo_{0.05}Al$ and $Fe_2Cr_{0.85}Mo_{0.15}Al$, $\log_{10} \chi^{-1}$ Vs $\log_{10}$ (T/$T_C^R$-1) is plotted (shown in inset of fig. 7 (a)-(b)) and slopes obtained from fitted with eqn. (6) in both PM and GP region are $\lambda_P$ and $\lambda_G$ respectively. The obtained values of $T_C^R$, $\lambda_P$ and $\lambda_G$ are given in table 5. In $Fe_2Cr_{0.95}Mo_{0.05}Al$, value of $\lambda_G$ is found to be greater than obtained in $Fe_2CrAl$, which further increases with higher concentration of Mo. This indicates that with an increase in Mo concentration GP becomes stronger and shifts towards high temperature side. This is expected as the anti-site disorder responsible for the observation of GP, increases with Mo concentration.

### 3.8 Magnetocaloric studies

Interestingly, from *M* (*H*) isotherms we have noted that these alloys exhibit soft FM behaviour which indicates that energy loss will be minimal during field cycle. Hence, we have investigated the magnetocaloric properties of these alloys. For this purpose, we have measured field dependent magnetization *M* (*H*) data for $Fe_2Cr_{1-x}Mo_xAl$ (x = 0, 0.05 and 0.15) alloys across the $T_C$. We have calculated the MCE in term of isothermal magnetic entropy change ($\Delta S_M$) produced by the magnetic field change. Using the obtained virgin *M* (*H*) curves, we have determined $\Delta S_M$ using the equation [47]:

$$\Delta S_M = \sum \frac{M_{n+1} - M_n}{T_{n+1} - T_n} \Delta H_n \ldots\ldots\ldots\ldots (7)$$

where $M_{n+1}$ and $M_n$ are the values of magnetization obtained at temperature $T_{n+1}$ and $T_n$ respectively. Fig. 8 (a-c) shows the temperature dependent $\Delta S_M$ curve at different $\Delta H$ for $Fe_2Cr_{1-x}Mo_xAl$ alloys. All the alloys show a peak in $\Delta S_M$ near $T_C$. It is noted that there is partial increment of $\Delta S_M$ with Mo concentration. The calculated value of $\Delta S_M$ for $Fe_2CrAl$, $FeCr_{0.95}Mo_{0.05}Al$ and $Fe_2Cr_{0.85}Mo_{0.15}Al$ are 1.19 J/kg-K, 1.60 J/Kg-K and 1.82 J/kg-K respectively at $\Delta H$ = 70 kOe. The value of $\Delta S_M$ is lower than those noted in materials which show significant magnetocaloric effect. Generally, materials which undergo second order phase transition (SOPT), exhibit small magnetic and thermal loss. To determine the nature of transition in this series of alloys we followed a method suggested by Franco and Conde [49]. According to the method, if the nature of phase transition is of second order then after scaling all the entropy curves obtained at different applied magnetic fields should collapse into a single curve [48]. To obtain this, entropy should be normalized, and temperature should be scaled into $t$ using [49]:

$$t = -\frac{(T-T_{pk})}{(T_{R1}-T_{pk})}, \quad T \leq T_C \quad \ldots\ldots (8)$$

$$\frac{(T-T_{pk})}{(T_{R2}-T_{pk})}, \quad T > T_C \ldots\ldots (9)$$

where $T_{R1}$ and $T_{R2}$ are the reference temperature below and above $T_C$ respectively, $T_{pk}$ is the temperature corresponding to peak temperature of $\Delta S_M$. Here we have used normalized entropy as $\Delta S_M /\Delta S_M^{max} = 0.5$. Fig. 8 (d-f) shows the normalized $\Delta S_M$ curves at different $\Delta H$ merging into single curve which is signature of SOPT in $Fe_2Cr_{1-x}Mo_xAl$ (x = 0, 0.05 and 0.15). We have further analyzed MCE in these alloys using mean field theory, which tells us the dependency of $\Delta S_M$ with applied field [47]. Here, $\Delta S_M$ show power law dependency with $H$ as $\Delta S_M = a\,(H)^n$ where $a$ is the constant and $n$ is the exponent describing the magnetic state of the system. Generally, the value of $n$ is equal to 1 and 2 at temperature below and above $T_C$ respectively. This signifies that across the transition the value of $n$ should change from 1 to 2 i.e. $n = 1$ in FM region and attains minimum value at $T_C$ whereas $n = 2$ in the PM region. This type of dependency has been earlier explored in different magnetic systems [49]. Fig. 8 (g-i) shows the temperature dependent behaviour of $n$. In $Fe_2CrAl$, the value of $n$ approaches 1 below $T_C$ while above $T_C$ it is less than 2. Similarly, in $Fe_2Cr_{0.95}Mo_{0.05}Al$ and $Fe_2Cr_{0.85}Mo_{0.15}Al$, below $T_C$ $n$ approaches 1 whereas above $T_C$ it is less than 2. This indicates the presence of some short-range correlations in PM region of both alloys, which is in accordance with presence of GP above $T_C$.

**Discussions**

Our systematic investigation of $FeCr_{0.95}Mo_{0.05}Al$ and $Fe_2Cr_{0.95}Mo_{0.15}Al$ reveal that Mo substitution at Cr-site results in decrement of $T_C$ in comparison to $Fe_2CrAl$. This observation is due to the effect of a strong hybridization between 3d-3d states of Fe-Cr elements and 3d-4d states of Fe-Mo elements, which increases with Mo concentration. In $Fe_{2-x}Mn_xCrAl$ ($0 \leq x \leq 1$), similar trend of $T_C$ was observed due to incorporation of AFM-Mn at FM-Fe site. Additionally, a shift in $T_f$ towards lower temperature is noted with increment in Mo substitution which can be ascribed as the effect of reduction in strength of magnetic anisotropy. While in Fe$_{2-x}$Mn$_x$CrAl, $T_f$ gets shifted towards higher temperature as the strength of magnetic anisotropy increases with Mn concentration [20]. In high temperature regime (above $T_C$), an increasing trend in the value of $\lambda_G$ in GP is noted. This indicates strengthening of GP with Mo substitution due to anti-site disorder. It results in the formation of the highly inhomogeneous phase leading to coexistence of short-range magnetic correlations.

## 4. Conclusions

In conclusion, we have investigated effect of partial substitution of Mo at Cr site on the electronic structure and physical properties of Fe$_2$CrAl. The XRD and high-resolution SEM data confirm perfect crystalline single phase of Fe$_2$Cr$_{0.95}$Mo$_{0.05}$Al and Fe$_2$Cr$_{0.85}$Cr$_{0.15}$Al alloys. However, the absence of (200) and (220) peaks indicates the presence of anti-site disorder in both alloys. With increment in Mo concentration, shift in Griffiths phase towards higher temperature as compared to Fe$_2$CrAl is observed. The formation of GP is attributed to presence of anti-site disorder. Moreover, an decrement in $T_C$ with respect to parent alloy is noted with increasing Mo substitution, which can be understood as the affect of increased hybridization strength between 3$d$-4$d$ states of Fe/Cr/Mo. Interestingly, similar to parent alloy cluster glass phase persists in Fe$_2$Cr$_{0.95}$Mo$_{0.05}$Al, which arises due to the effect of magnetic anisotropy. Additionally, our studies may provide a pathway to explore the effect of atomic disorder on the electronic structure and physical properties of transition metal substituted Fe$_2$CrAl, through microscopic probes and computational tools.

**Acknowledgements**

K.M. acknowledges the financial support from a research grant (Grant No. 03(1381)/16/EMR-II) from SERB, India. The authors acknowledge the experimental facilities of Advanced Materials Research Center (AMRC), IIT Mandi.

**References**

1. Klimczuk T., Wang C. H., Gofryk K., Ronning F., Winterlik J., Fecher G. H., Griveau J. C., Colineau E., C. Felser C., Thompson J. D., Safarik D. J., and Cava R. J. 2012 Superconductivity in the Heusler family of intermetallics *Phys. Rev. B* **85** 174505
2. de Groot R. A. and Buschow K. H. J. 1986 Recent developments in Half-metallic magnetism *J. Magn. Magn. Mater.* **54** 1377

3. van Engen P. G., Buschow K. H. J., Jongebreur R. and Emran R. 1983 PtMnSb, a material with very high magnetooptical Kerr effect *Appl. Phys. Lett.* **42** 202

4. Krenke T., Duman E., Acet M., Wassermann E. F., Moya X., Mañosa L., and Planes A. Inverse magnetocaloric effect in ferromagnetic Ni-Mn-Sn alloys 2005 *Nat. Mat.* **4** 450

5. Zhang W., Sun Y., Wang H., Li Y., Zhang X., Sui Y., Luo H., Meng F., Qian Z., and Wu G. 2014 The spin glass behavior in the Heusler alloy $Cu_2VAl$ *Journal of alloys and compds.* **589** 230

6. de Groot R. A., Mueller F. M., van Engen P. G., and Buschow K. H. J. 1983 New class of Materials: Half-metallic ferromagnets *Phys. Rev. Lett.* **50** 2024

7. Pickett W.E. and Moodera J. S. 2001 Half metallic magnets *Phys. Today* **54** 39

8. Zutic I., Fabian J., and Sharma S. D. 2014 Spintronics: Fundamentals and applications *Rev. Mod. Phys.* **76** 323

9. Basso V., Sasso C. P., Skokov K. P., Gutfleisch O., and Khovaylo V. V. 2012 Hystersis and magnetocaloric effect at the magnetostructural phase transition of Ni-Mn-Ga and Ni-Mn-Co-Sn Heusler alloys 2012 *Phys. Rev. B* **85** 014430

10. Sahoo R., Nayak A. K., Suresh K. G., and Nigam A. K. 2011 In-plane and out of plane magnetic properties in $Ni_{46}Co_4Mn_{38}Sb_{12}$ ribbons *J. Appl. Phys.* **109** 07A921

11. Dubenko I., Pathak A., Stadler S., Ali N., Kovarskii Y., Prudnikov V. N., Perov N. S., and Granovsky A. B. 2009 Giant Hall effect in N-Mn-In Heusler alloys *Phys. Rev. B*, 2009 092408(R)

12. Dubenko I., Ali N., Stadler S., Zhukov A., Zhukova V., Hernando B., Prida V., Prudnikov V., Gan'shina E., and Granovsky A., *Novel Functional Magnetic Materials: Fundamentals and Applications*, Springer Series in Materials Science Vol. 231, Chap. II, edited by A. Zhukov (Springer, 2016), p. 41.

13. Slebraski A., Wrona A., T. Zawada T., Jezierski A., Zygmunt A., Szot K., Chiuzbaian S., and Neumann M. 2002 Electronic structure of some Heusler alloys based on aluminium and tin *Phys. Rev. B* **65** 144430

14. Kawakami M. 1993 Huge increase in the magnetic ordering temperature of $Fe_2Mn_{1-x}V_xSi$ at intermediate composition *J. Magn. Magn. Mater.* **128** 284

15. Guo G. Y., Botton G. A., and Nishino Y. 1998 Electronic structure of possible 3d 'Heavy fermion' compound $Fe_2VAl$ J. *Phys.: Condens. Matter* **10** L119

16. Weht R., and Pickett W. E. 1998 Excitonic correlations in the intermetallic $Fe_2VAl$ *Phys. Rev. B* **58** 6855

17. Bansil A., Kaprzyk S., Mijnarends P. E., and Tobola J. 1999 Electronic structure and magnetism of $Fe_{3-x}V_xX$ (x=Si, Ga and Al) alloys by KKR-CPA method Phys. *Rev. B* **60** 13396

18. Singh D. J. and Mazin I. I. 1998 Electronic structure, local moments, and transport in $Fe_2VAl$ *Phys. Rev. B* **57** 14352

19. Nishino Y., Sumi H., and Mizutani U. 2005 Transport and magnetic properties of the Heusler –type $Fe_{2-x}V_{1+x}Al$ system (-0.01≤x≤ 0.08) *Phys. Rev. B* **71** 094425

20. Yadav K., Sharma M. K., Singh S., and Mukherjee K. 2019 Exotic magnetic behaviour and evidence of cluster glass and Griffiths like phase in Heusler alloys $Fe_{2-x}Mn_xCrAl$ (0≤x≤1) *Sci. Rep.* **9** 15888

21. Kellou A., Fenineche N. E., Grosdidier T., Aourag H., and Coddet C. 2003 Structural stability and magnetic properties in X2AlX′ (X=Fe, Co, Ni; X′=Ti, Cr) Heusler alloys from quantum mechanical caluclations *J. Appl. Phys.* **94** 3292

22. Saha R., Srinivas V., and Chandrasekhar T. V. Rao 2009 Evolution of ferromagnetic like order in $Fe_2V_{1-x}Cr_xAl$ Heusler alloys *Phys. Rev. B* **79** 174423

23. Venkateswarlu B., Babu P. D., and Kumar Narayanan H. 2014 Complex magnetic behavior of the Heusler alloy-$Cu_2Mn_{0.75}Fe_{0.25}Al$ *IEEE Transc. On Magnetics* **50** 11

24. Feng Y., Chen H., Yuan H., Zhou Y., and Chen X. 2015 Thermodynamic stability, magnetism and Half metallicity of Mn2CoAl/GaAs (001) Interface *J. Magn. Magn. Mater.* **378** 7

25. Borgohain P. and Sahariah M. B. 2015 Effect of compositional and antisite disorder on the electronic and magnetic properties of Ni-Mn-In Heusler alloy *J. Phys.: Condens. Matter* **27** 175502

26. Lin T. T., Dai X. F., Duo R. K., Cheng Z. X., Whang L. Y., Whang X. T., and Liu G. D. 2017 Anti-site induced diverse diluted magnetism in LiMgPdSb-type CoMnTiSi alloy *Sci. Rep.* **7** 42034

27. Tougaard S. 1989 Practical algorithm for background subtraction *Surf. Sci.* **216** 343-360

28. Nehla P., Ulrich C., and Dhaka R. S. 2019 Investigation of the structural, electronic, transport and magnetic properties of $Co_2FeGa$ Heusler alloys *J. Alloys and Cmpds.* **776** 379-386

29. Mehlhorn W. 1985 Auger-Electron Spectrometry of Core Levels of Atoms. In: B. Crasemann B. Atomic Inner-Shell Physics. Physics of Atoms and Molecules. Springer, Boston, MA

30. Slebarski A., Neumann M., and Schneider B. 2001 Magnetic splitting in x-ray photoelectron spectroscopy Cr L spectra of $Fe_2CrAl$, $Co_2CrAl$ and $Cu_2CrAl$ *J. Phys.: Condens. Matter* **13** 5515

31. Mydosh J. A. 1993 Spin Glasses: An experimental introduction, Taylor and Francis, London

32. Sharma M. K., Yadav K., and Mukherjee K. 2018 Complex magnetic behaviour and evidence of a superspin glass state in the binary intermetallic compound $Er_5Pd_2$ *J. Phys.: Condens. Matter* **30** 215803

33. Sharma M. K. and Mukherjee K. 2018 Evidence of large magnetic cooling power and double glass transition in $Tb_5Pd_2$ *J. Magn. Magn. Mater.* **466** 317

34. Galanakis I., Mavropoulos Ph., and Papanikolaou N. 2002 Slater- pauling Behvaior and origin of the half-metallicity of the full Heusler alloys *Phys. Rev. B* **66** 174429

35. Binder K. and Young A. P. 1986 Spin glasses: Experimental facts, theoretical concepts, and open questions *Rev. Mod. Phys.* **58** 801

36. Sun Y., Salamon M. B., Garnier K., and Averback R. S. 2003 Memory effects in an interacting magnetic nanoparticle systems *Phys. Rev. Letters* **91** 167206

37. Bandopadhya M. and Dasgupta S. 2006 Memory in nanoparticle systems: Superparamagnetism versus spin- glass behavior Phys. *Rev. B* **74** 214410

38. Pakhira S., Mazumdar C., Ranganathan R., Giri S., and Avdeev M. 2016 Large magnetic cooling power involving frustrated antiferromagnetic spin glass state in $R_2NiSi_3$ (R=Gd, Er) *Phys. Rev. B* 94 104414

39. Malinowski A., Bezusyy V., Minikayev R., Dziawa P., Syryanyy Y., and Sawicki M. 2011 Spin glass behvaior in Ni-doped $La_{1.85}Sr_{0.15}CuO_4$ *Phys. Rev. B*, 2011, **84** , 024409

40. Anand V. K., Adroja D.T., and Hillier A. D. 2012 Ferromagnetic cluster spin glass behavior in $PrRhSn_3$ *Phys. Rev. B* **85** 014418

41. Mukherjee K., and Banerjee A. 2009 Selective substitution in orbital domains of a low doped manganite: an investigation from Griffiths phenomenon and modification of glass features J. *Phys: Condens. Matter* **21** 106001

42. Salamon M. B., Lin P., and Chun S. H. 2002 Colossal Magnetoresistance is a Griffiths singularity *Phys. Rev. Lett.* **88** 197203
43. Deisenhofer J., Braak D., Krug von Nidda H.-A, Hemberger J., Eremina R. M., Ivanshin V. A., Balbashov A. M., Jug G., Loidl A., Kimura T., and Tokura Y. 2005 Observation of Griffiths phase in paramagnetic $La_{1-x}Sr_xMnO3$ *Phys. Rev. Lett.* **95** 257202
44. Magen C., Algarabel P. A., Morellon L., Araújo J. P., Ritter C., Ibarra M. R., Pereira A. M., and Sousa J. B. 2006 Observation of a Griffiths-like phase in the magnetocaloric compound $Sr_3CuRhO_3$ *Phys. Rev. Lett.* **96** 167201
45. Sampathkumaran E. V., Mohapatra N., Rayaprol S., and Iyer K. K. 2007 Magnetic anomalies in the spin chain compound $Sr_3CuRhO_3$: Griffiths-phase-like behavior of magnetic susceptibility *Phys. Rev. B* **75** 052412
46. He C., Torija M. A., Wu J., Lynn J. W., Zheng H., Mitchell J. F., and Leighton C. 2007 Non-Griffiths-like clustered phase above the Curie temperature of the doped perovskite cobaltite $La_{1-x}Sr_xCoO_3$ *Phys. Rev. B* **76** 014401
47. Pekala M. 2010 Magnetic field dependence of magnetic entropy change in nanocrystalline and polycrystalline manganites $La_{1-x}M_xMnO_3$ (M=Ca, Sr) *J. Appl. Phys.* **108** 113913
48. Bonilla C. M., Herrero-Albillos J., Bartolome F., Gracia L. M., Parra-Borderias M., and Franco V. 2010 Universal behavior for magnetic entropy change in magnetocaloric materials: An analysis on the nature of phase transitions *Phys. Rev. B* **81** 224424
49. Franco V., Conde A., Romero-Enrique J. M., and Blazquez J. S. 2008 A universal curve for the magnetocaloric effect: an analysis based on scaling relations *J. Phys.: Condens. Matter* **20** 285207

**Tables-**

Table 1: Obtained lattice parameters and unit cell volume of $Fe_2Cr_{1-x}Mo_xAl$ (x = 0, 0.05 and 0.15)

|  | Lattice parameter (Å) | Unit cell volume (Å$^3$) |  |
|---|---|---|---|
| $Fe_2CrAl$ | 5.784±0.001 | 193.383±0.001 | Ref. [20] |
| $Fe_2Cr_{0.95}Mo_{0.05}Al$ | 5.792±0.06 | 194.300±0.001 | This work |
| $Fe_2Cr_{0.85}Mo_{0.15}Al$ | 5.81±0.09 | 196.120±0.001 | This work |

Table 2: Binding energy of $2p_{3/2}$, $2p_{1/2}$, $2p$, $3p_{3/2}$ and $3p_{1/2}$ levels of respective alloys

|  | Fe (eV) | Fe (eV) | Cr (eV) | Cr (eV) | Al (eV) | Mo (eV) | Mo (eV) |
|---|---|---|---|---|---|---|---|
|  | $2p_{3/2}$ | $2p_{1/2}$ | $2p_{3/2}$ | $2p_{1/2}$ | $2p$ | $3p_{3/2}$ | $3p_{1/2}$ |
| $Fe_2CrAl$ | 707.30 | 720.21 | 574.46 | 583.63 | 72.67 | - | - |
| $Fe_2Cr_{0.95}Mo_{0.05}Al$ | 707.37 | 720.30 | 574.56 | 583.83 | 72.8 | 394.13 | 411.81 |
| $Fe_2Cr_{0.85}Mo_{0.15}Al$ | 707.42 | 720.42 | 574.86 | 583.90 | 73.5 | 394.29 | 411.90 |

Table 3: Parameters obtained from fitting of eqn. 1 in *M (H)* data at 2 K

|  | $M_0$ (emu/gm) | $H_{ex}$ (kOe) | $H_r$ (kOe) |  |
|---|---|---|---|---|
| $Fe_2CrAl$ | 46.3±0.07 | 29.1±1.9 | 29.0±1.8 | Ref.20 |
| $Fe_2Cr_{0.95}Mo_{0.05}Al$ | 46.45±0.08 | 28.73±3.72 | 27.80±3.44 | This work |
| $Fe_2Cr_{0.95}Mo_{0.05}Al$ | 46.66±0.07 | 28.42±4.21 | 25.05±3.35 | This work |

Table 4: Parameters obtained from fitting of time dependent magnetization data with eqn. 2

|  | T (K) | $M_0$ (emu/gm) | S (emu/gm) |
|---|---|---|---|
| $Fe_2Cr_{0.95}Mo_{0.05}Al$ | 2 | 0.362±0.001 | 0.0061±0.0002 |
|  | 4 | 0.167±0.002 | 0.0003±0.0001 |
| $Fe_2Cr_{0.85}Mo_{0.15}Al$ | 2 | 0.515±0.001 | 0.0007±0.0002 |

Table 5: Parameters obtained from straight line fitting of log-log plot of $\chi_{DC}^{-1}$ as function of $(T/T_C^R)$-1 in PM and GP region

|  | $T_C^R$ (K) | $\lambda_P$ | $\lambda_{GP}$ |  |
|---|---|---|---|---|
| $Fe_2CrAl$ | 313 | 0.0004 | 0.84±0.02 | Ref.20 |
| $Fe_2Cr_{0.95}Mo_{0.05}Al$ | 323.4 | 0.0094 | 0.85±0.03 | This work |
| $Fe_2Cr_{0.85}Mo_{0.15}Al$ | 338 | 0.0054 | 0.89±0.05 | This work |

**Figures-**

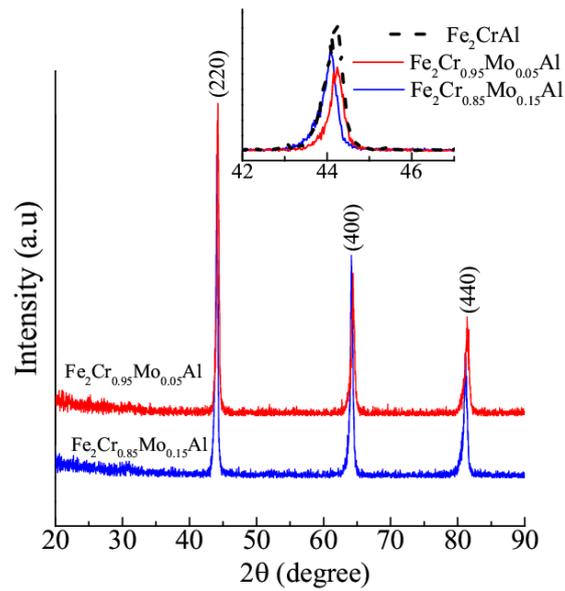

Figure 1 Room temperature XRD patterns of annealed $Fe_2Cr_{1-x}Mo_xAl$ (x = 0.05 and 0.15) alloys Inset: shows the shifting of peak with increasing Mo substitution of $Fe_2CrAl$, $Fe_2Cr_{0.95}Mo_{0.05}Al$ and $Fe_2Cr_{0.85}Mo_{0.15}Al$; XRD pattern of $Fe_2CrAl$ is taken from Ref. [20] for comparison.

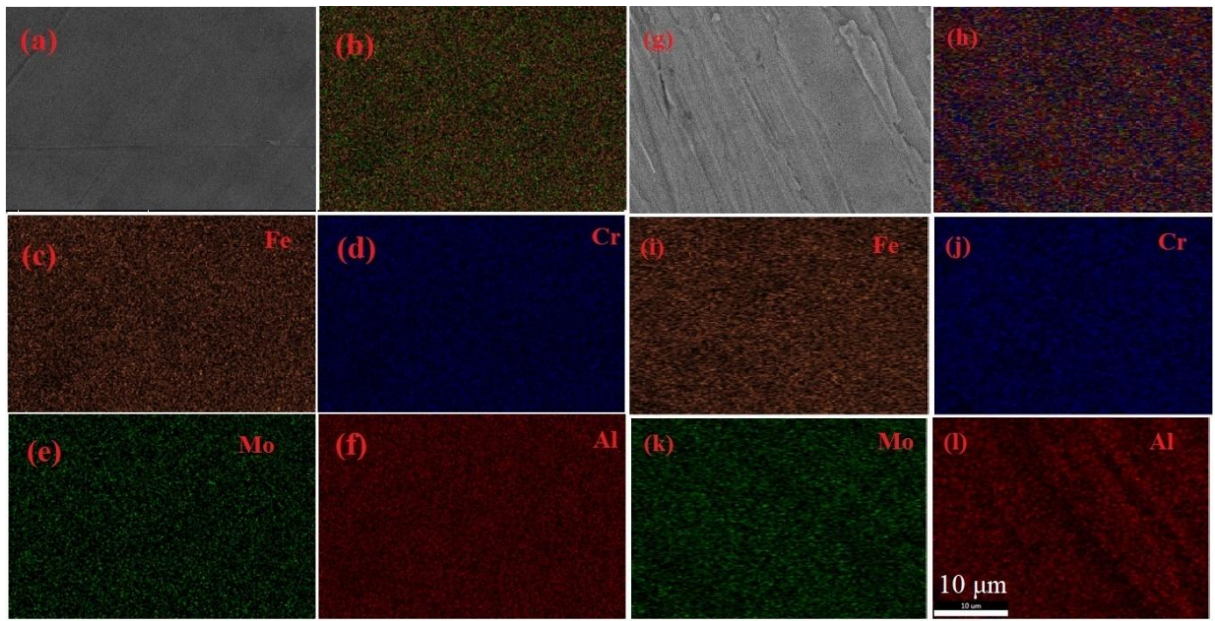

Figure 2 (a) and (g) represents the SEM image of $Fe_2Cr_{0.95}Mo_{0.05}Al$ and $Fe_2Cr_{0.85}Mo_{0.15}Al$ (b) and (h) depict all together distribution for Fe, Cr, Mo and Al in $Fe_2Cr_{0.95}Mo_{0.05}Al$ and $Fe_2Cr_{0.85}Mo_{0.15}Al$ (c)-(f) EDS elemental mapping for Fe, Cr, Mo and Al in $Fe_2Cr_{0.95}Mo_{0.05}Al$ (i)-(l) EDS elemental mapping for Fe, Cr, Mo and Al in $Fe_2Cr_{0.85}Mo_{0.15}Al$. Scale bars are 10 μm

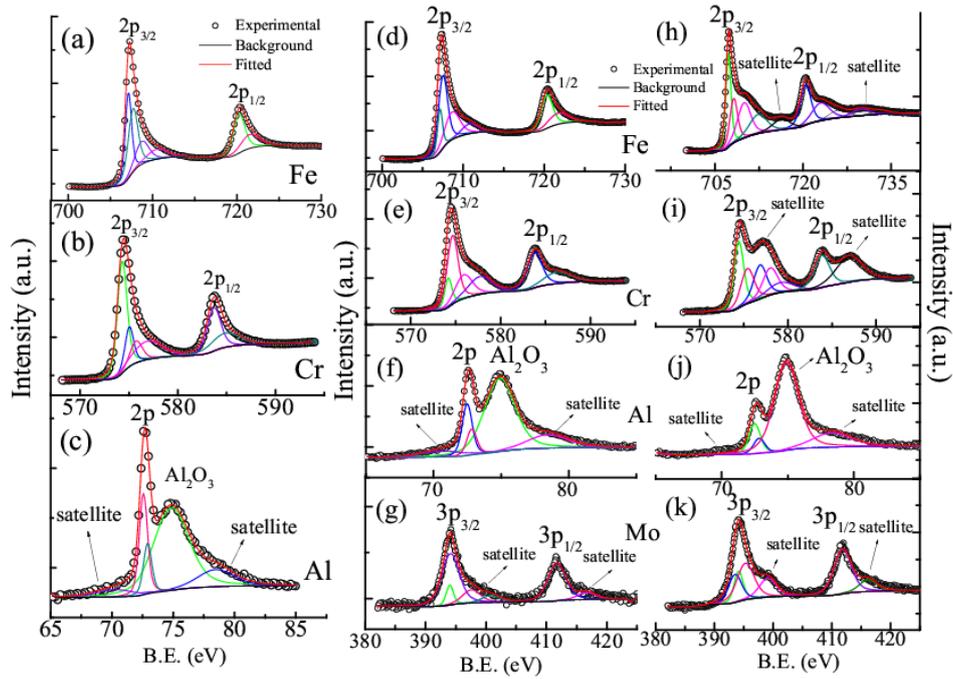

Figure 3 Core level XPS spectra of respective elements of Fe$_2$CrAl (left panel), Fe$_2$Cr$_{0.95}$Mo$_{0.05}$Al (middle panel) and Fe$_2$Cr$_{0.85}$Mo$_{0.15}$Al (right panel), respectively.

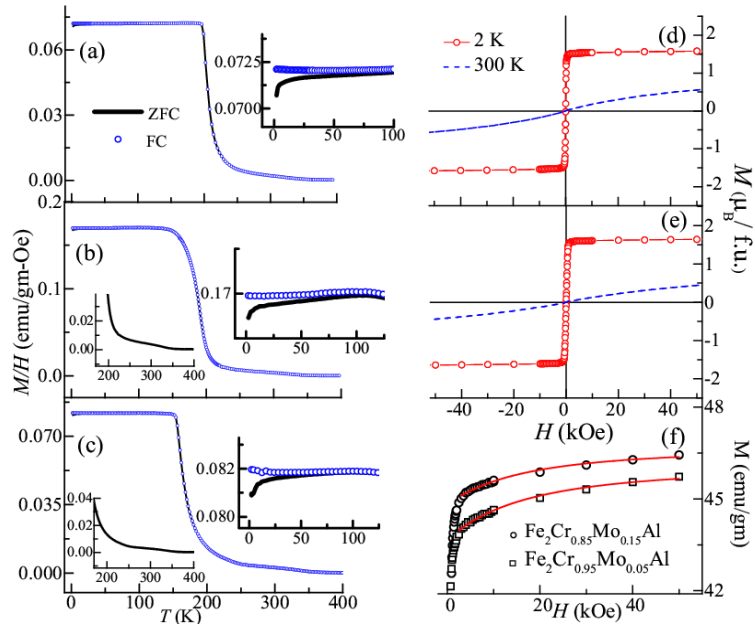

Figure 4 Temperature response of DC susceptibility of (a) Fe$_2$CrAl [from Ref. 20], (b) Fe$_2$Cr$_{0.95}$Mo$_{0.05}$Al, and (c) Fe$_2$Cr$_{0.85}$Mo$_{0.15}$Al obtained under ZFC and FC protocol at 100 Oe in 2-400 K temperature range. Left insets of (b)-(c): magnified view of ZFC plot in the temperature range 170-400 K. Right insets of (a)-(c) magnified view of ZFC and FC plot of both alloys below 125 K. Field dependent behaviour of magnetization at 2 and 300 K of (d) Fe$_2$Cr$_{0.95}$Mo$_{0.05}$Al and (e) Fe$_2$Cr$_{0.85}$Mo$_{0.15}$Al (f) Virgin $M(H)$ of both alloys at 2 K with fit (red solid line) using eqn. 1.

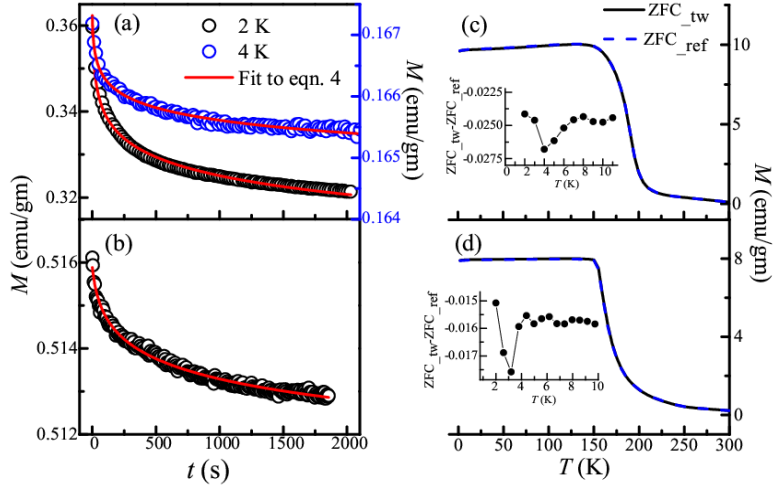

Figure 5 Time dependent relaxation behaviour of magnetization at 2 and 4 K of (a) $Fe_2Cr_{0.95}Mo_{0.05}Al$ and (b) $Fe_2Cr_{0.85}Mo_{0.15}Al$; Temperature dependent ZFC memory effect of (c) $Fe_2Cr_{0.95}Mo_{0.05}Al$ and (d) $Fe_2Cr_{0.85}Mo_{0.15}Al$ at different waiting temperatures Inset: Temperature response of (ZFC$_{\_tw}$-ZFC$_{\_ref}$) in the low temperature range.

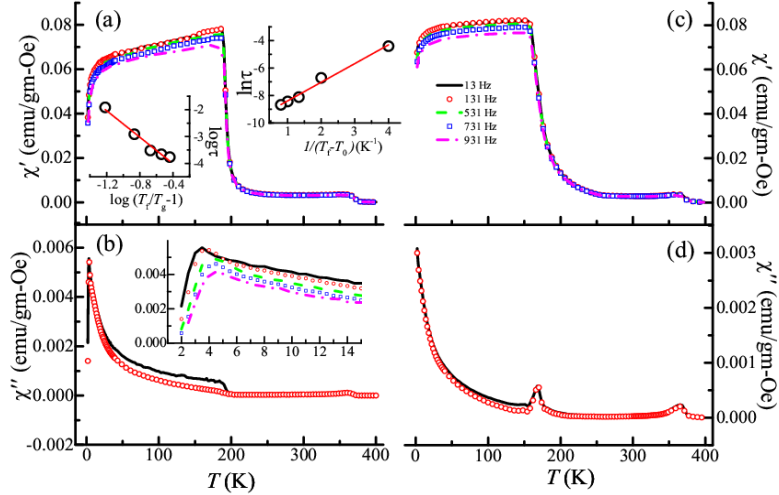

Figure 6 (a) and (c) Temperature response of in-phase ($\chi'$) of AC susceptibility measured at different frequencies for $Fe_2Cr_{0.95}Mo_{0.05}Al$ and $Fe_2Cr_{0.85}Mo_{0.15}Al$ respectively. (b) and (d) Temperature response of out-of-phase ($\chi''$) of AC susceptibility measured at different frequencies for $Fe_2Cr_{0.95}Mo_{0.05}Al$ and $Fe_2Cr_{0.85}Mo_{0.15}Al$ (only two frequencies are shown for clarity). Left inset of (a) Critical law fit of relaxation time ($\tau$) as function of reduced temperature ($T_f/T_g$-1) using eqn. 4. Right inset of (a) V-F law fit of relaxation time ($\tau$) as function of reduced temperature $1/(T_f-T_0)$ using eqn. 5. Inset of (b): $\chi''$ plot at different frequencies (13-931 Hz) in the temperature range 1.8-15 K to show the shift in peak temperature.

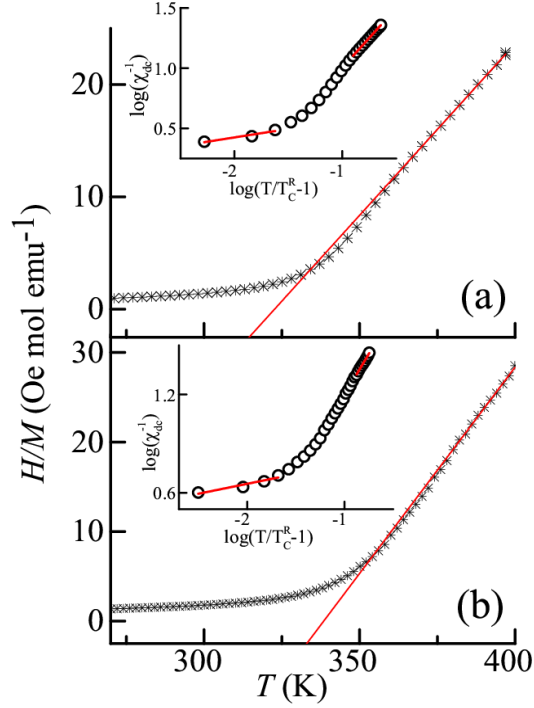

Figure 7 Inverse DC susceptibility as function of temperature plot at 100 Oe of (a) $Fe_2Cr_{0.95}Mo_{0.05}Al$ and (b) $Fe_2Cr_{0.85}Mo_{0.15}Al$. Solid red line represents the fitting using CW law. Insets represent the log-log plot of $\chi^{-1}_{DC}$ vs. $T/T_C^R - 1$. Solid red lines are straight line fitting in GP and PM region.

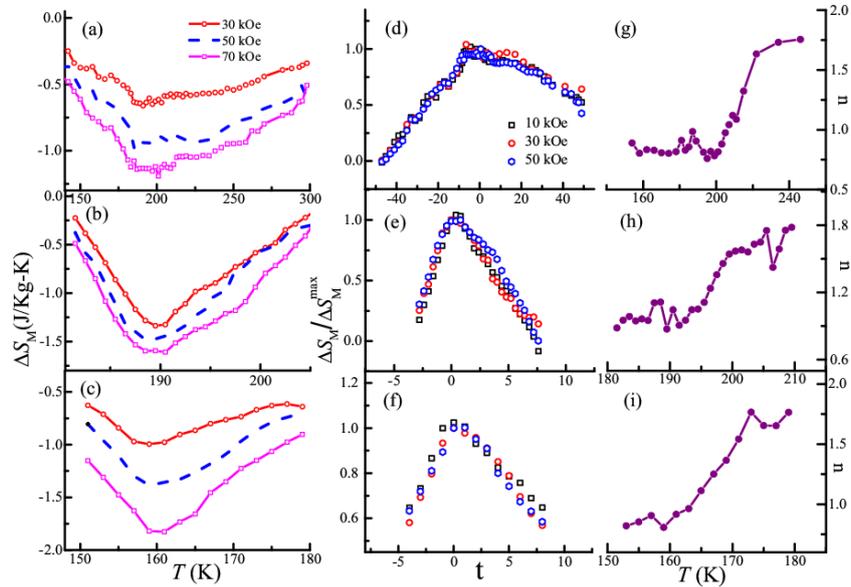

Figure 8 Left panel: Temperature response of $\Delta S_M$ of (a) $Fe_2CrAl$ (b) $Fe_2Cr_{0.95}Mo_{0.05}Al$ and (c) $Fe_2Cr_{0.85}Mo_{0.15}Al$ at different $\Delta H$; Middle panel: $\Delta S_M / \Delta S_M^{max}$ as a function of reduced temperature (t) plot of (d) $Fe_2CrAl$ (e) $Fe_2Cr_{0.95}Mo_{0.05}Al$ and (f) $Fe_2Cr_{0.85}Mo_{0.15}Al$; Right panel: plot showing variation of value of $n$ with temperature for (g) $Fe_2CrAl$ (h) $Fe_2Cr_{0.95}Mo_{0.05}Al$ and (i) $Fe_2Cr_{0.85}Mo_{0.15}Al$